\documentclass[astronomy,article,accept,moreauthors,pdftex]{Definitions/mdpi}

\usepackage{gensymb}

\firstpage{1} 
\pubvolume{1}
\issuenum{1}
\articlenumber{0}
\pubyear{2022}
\copyrightyear{2022}
\externaleditor{Academic Editor: Ignatios Antoniadis} 
\datereceived{13 May 2022} 
\dateaccepted{27 July 2022} 
\datepublished{2 August 2022} 
\hreflink{https://doi.org/} 

\Title{Rotating and Expanding Gas in~Binary Post-AGB Stars}

\TitleCitation{Rotating and Expanding Gas in~Binary Post-AGB Stars}

\Author{Iván Gallardo Cava$^{1,}$*, Valentín Bujarrabal $^{1}$, Javier Alcolea $^{1}$, Miguel Gómez-Garrido $^{1,2}$, Arancha~Castro-Carrizo $^{3}$, Hans Van Winckel $^{4}$ and Miguel Santander-García $^{1}$}

\AuthorNames{Iván Gallardo Cava, Valentín Bujarrabal, Javier Alcolea, Miguel Gómez-Garrido, Arancha Castro-Carrizo, Hans Van Winckel and Miguel Santander-García}

\AuthorCitation{Gallardo Cava, I.;Bujarrabal, V.; Alcolea, J.; Gómez-Garrido, M.; Castro-Carrizo, A.; Van Winckel, H.; Santander-García, M.}

\address{%
$^{1}$ \quad Observatorio Astronómico Nacional (OAN-IGN), Alfonso XII 3, 28014 Madrid, Spain; v.bujarrabal@oan.es~(V.B.); j.alcolea@oan.es (J.A.); m.gomezgarrido@oan.es (M.G.-G.); m.santander@oan.es~(M.S.-G.)\\
$^{2}$ \quad Centro de Desarrollos Tecnológicos, Observatorio de Yebes (IGN), 19141 Yebes, Spain\\ 
$^{3}$ \quad Institut de Radioastronomie Millimétrique, 300 rue de la Piscine, 38406 Saint-Martin-d’Hères, France; ccarrizo@iram.fr \\
$^{4}$ \quad Instituut voor Sterrenkunde, KU Leuven, Celestijnenlaan 200B, 3001 Leuven, Belgium; hans.vanwinckel@kuleuven.be}

\corres{Correspondence: i.gallardocava@oan.es}

\abstract{
There is a~class of binary post-AGB stars (binary system including a~post-AGB star) that are surrounded by Keplerian disks and outflows resulting from gas escaping from the~disk. To~date, there are seven sources that have been studied in~detail through interferometric millimeter-wave maps of CO lines (ALMA/NOEMA).
For the~cases of the~Red\,Rectangle, IW\,Carinae, IRAS\,08544-4431, and~AC\,Herculis, it~is found that around $\ge$85\% of the~total nebular mass is located in~the~disk with Keplerian dynamics. The~remainder of the~nebular mass is located in~an expanding component. This~outflow is probably a~disk wind consisting of material escaping from the~rotating disk. These~sources are the~disk-dominated nebulae. On the~contrary, our maps and modeling of 89\,Herculis, IRAS\,19125+0343, and~R\,Scuti, which allowed us to study their morphology, kinematics, and~mass distribution, suggest that, in~these sources, the~outflow clearly is the~dominant component of the~nebula ($\sim$75\% of the~total nebular mass), resulting in~a~new subclass of nebulae around binary post-AGB stars: the~outflow-dominated sources.
\\ Besides CO, the~chemistry of this type of source has been practically unknown thus far. We~also present a~very deep single-dish radio molecular survey in~the~1.3, 2, 3, 7, and~13\,mm bands ($\sim$600 h of telescope time). Our results and detections allow us to classify our sources as O- or /C-rich. We~also conclude that the~calculated abundances of the~detected molecular species other than CO are particularly low, compared with AGB stars. This~fact is very significant in~those sources where the~rotating disk is the~dominant component of the~nebula.} 


\begin{document}
\newcommand{\on}{89\,Her }
\newcommand{\onp}{89\,Her}
\newcommand{\iras}{IRAS\,19125+0343 }
\newcommand{\irasp}{IRAS\,19125+0343}
\newcommand{\ac}{AC\,Her }
\newcommand{\acp}{AC\,Her}
\newcommand{\rs}{R\,Sct }
\newcommand{\rsp}{R\,Sct}
\newcommand{\x}{\,$\times$\,}

\newcommand{\secp}{\mbox{\rlap{.}$''$}}



\section{Introduction}

There is a~type of post-AGB star characterized by their spectral energy distributions (SEDs), which shows a~near-infrared (NIR) excess indicating the~presence of hot dust close to the~the stellar system \citep[][]{vanwinckel2003,oomen2018}.
Their IR spectra reveal the~presence of highly processed dust grains, so~the dust might be located in~stable structures \citep[][]{gielen2011a,jura2003,sahai2011}.
All the above suggests the~presence of circumbinary disks.
Their disk-like shape has been confirmed by interferometric IR data see, e.g., \citep{hillen2017,kluska2019}). Their radial velocity curves reveal that the~post-AGB stars are part of a~binary system (see, e.g., \citep{oomen2018}). The~systematic detection of binary systems in~these objects strongly suggests that the~angular momentum of the~disks comes from the~stellar~system.

Observations of $^{12}$CO and $^{13}$CO in~the~$J= 2-1$ and $J= 1-0$ lines (230.538 and 220.398\,GHz, respectively) have been well analyzed in~sources with such~NIR excess \citep[][]{bujarrabal2013a}. There are two types of CO line profiles: 
(a) narrow CO line profiles characteristic of rotating disks and weak wings, which implies that most of the~nebular mass is contained in~the~disk (with Keplerian dynamics), and~(b) composite CO line profiles including a~narrow component, which very probably represents emission from the~rotating disk, and~strong wings, which represents emission from the~outflow, which could dominate the~nebula \citep[][]{bujarrabal2013a}.
These types of line profiles are also found in~young stars surrounded by a~rotating disk made of remnants of interstellar medium (ISM) and those expected from disk-emission modeling (see, e.g., \citep{bujarrabal2013a,guilloteau2013}). These~results indicate that the~CO emission lines of our sources come from disks with Keplerian or quasi-Keplerian~rotation.

The study of the~chemistry of this class of binary post-AGB stars, together with the~very detailed kinematic analysis of Keplerian disks and outflows around these sources, is based on published articles \mbox{(see \citep{gallardocava2021,gallardocava2022}).}

This paper is organized as follows. Technical information of our observations is given in~Section~\ref{observaciones}. In~Section~\ref{obs}, we~present mm-wave interferometric maps and models of the~most representative cases of a~disk-dominated nebula (AC\,Herculis), an~outflow-dominated nebula (R\,Scuti), and~an intermediate case in~between the~disk- and the~outflow-dominated nebula (89\,Herculis).
We present the~first molecular survey in~this kind of object in~Section~\ref{mole}, together with discussions about molecular intensities and chemistry. Finally, we~summarize our conclusions in~Section~\ref{sec5}.

\section{Observations} \label{observaciones}

We show interferometric maps of our sources using the~NOEMA interferometer. Observations of the~$^{12}$CO $J= 2-1$ rotational transition were carried out towards AC\,Herculis, R Scuti, and~89\,Herculis. Observations of the~$^{13}$CO $J= 2-1$ rotational transition were also obtained for 89\,Herculis. 

Our single-dish observations were performed using the~30\,m\,IRAM telescope \linebreak (Granada, Spain) and the~40\,m\,Yebes telescope (Guadalajara, Spain). We~observed at the~1.3, 2, 3, 7, and~13\,mm bands. Our observations required a~total telescope time of $\sim$600\,h distributed over the~two telescopes and for several projects to observe the~nebulae around the~next binary post-AGB stars: AC\,Her, the~Red\,Rectangle, HD\,52961, IRAS\,19157$-$0257, IRAS\,18123+0511, IRAS\,19125+0343, AI\,CMi, IRAS\,20056+1834, and~R\,Sct.


\section{NOEMA~Observations} \label{obs}

\begin{figure}[H]
\includegraphics[scale=0.40]{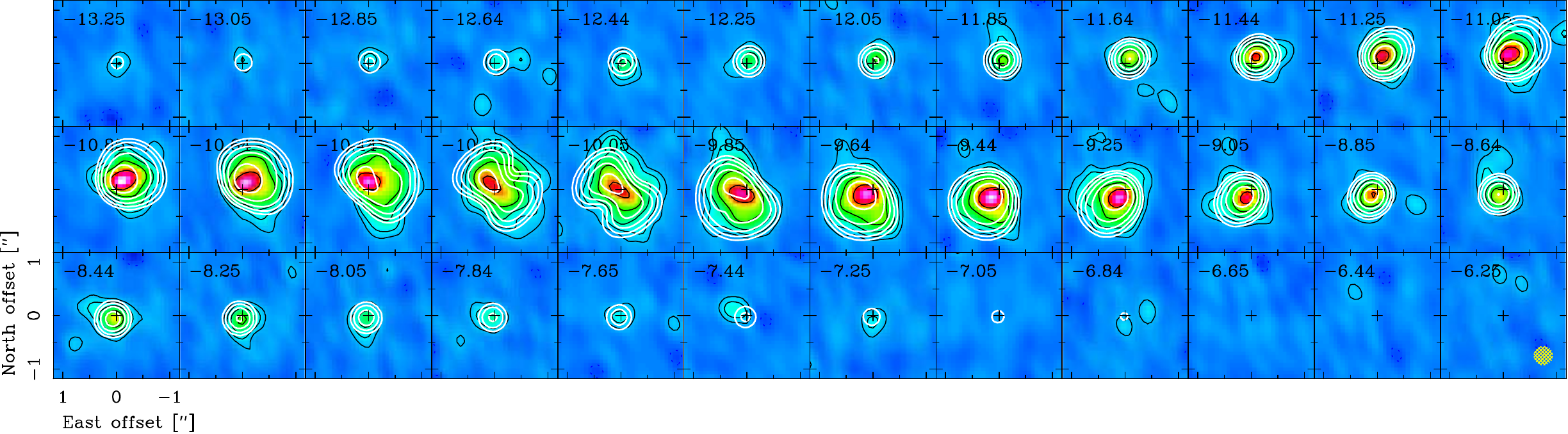}
 
\caption{NOEMA maps per velocity channel of AC\,Her in~$^{12}$CO $J=2-1$ emission. The~contours are $\pm$\,9, 18, 36, 76, and~144\,mJy\,beam$^{-1}$ with a~maximum emission of 230\,mJy\,beam$^{-1}$. The~LSR velocities are indicated in~each panel (upper-left corner) and the~beam size, 0.''35 \x 0.''35, is shown in~the~last panel at the~bottom right corner (yellow ellipse). 
We also show the~synthetic maps from our best-fit model in~white contours, to~be compared with observational data; the~scales and contours are the~same.}
    \label{fig:ac12mapas}  
\end{figure}

\begin{figure}[H]
        \begin{minipage}[b]{0.48\linewidth}
                \includegraphics[scale=0.25]{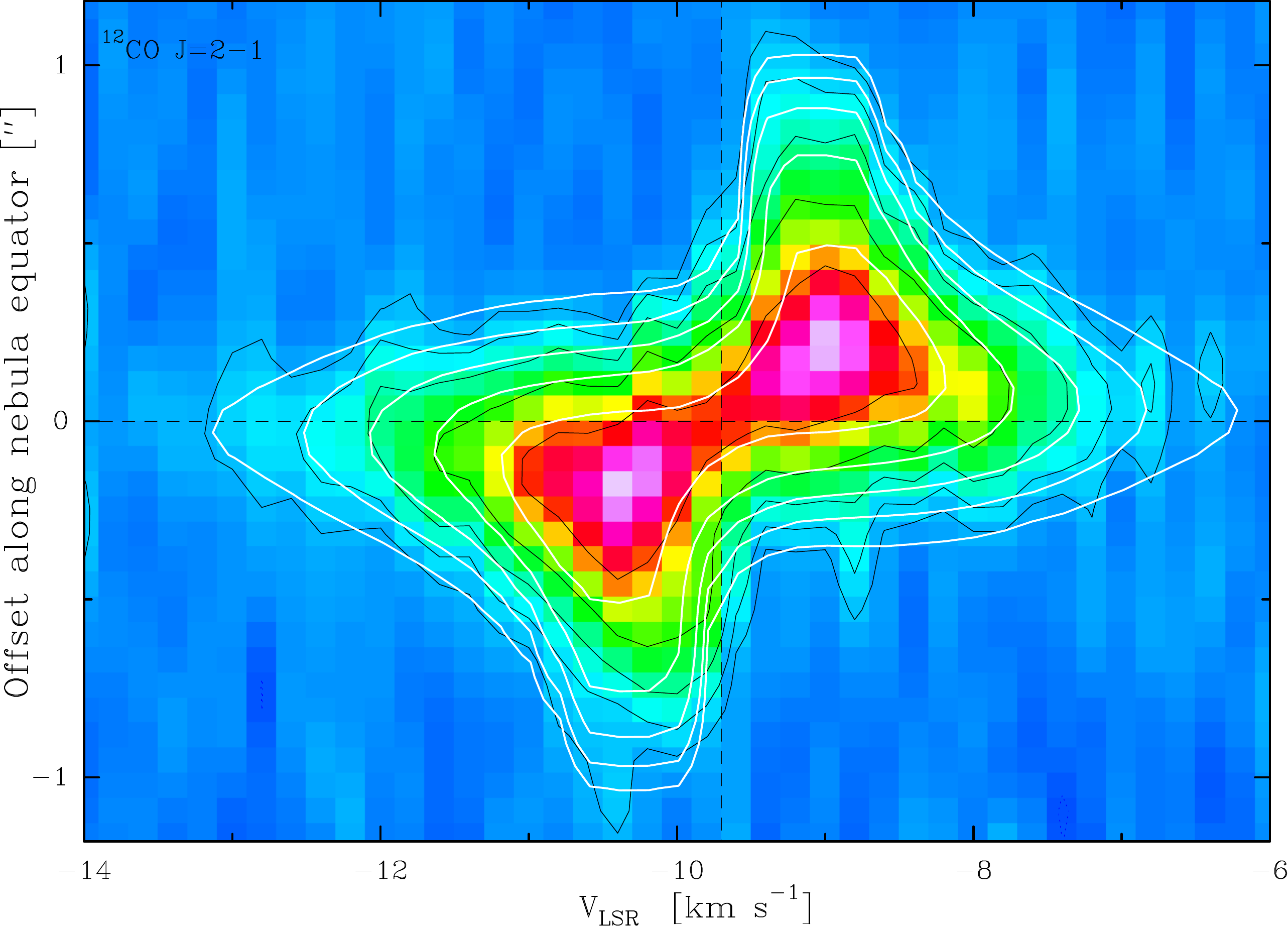}
        \end{minipage}
        \quad
        \begin{minipage}[b]{0.48\linewidth}
                \includegraphics[scale=0.25]{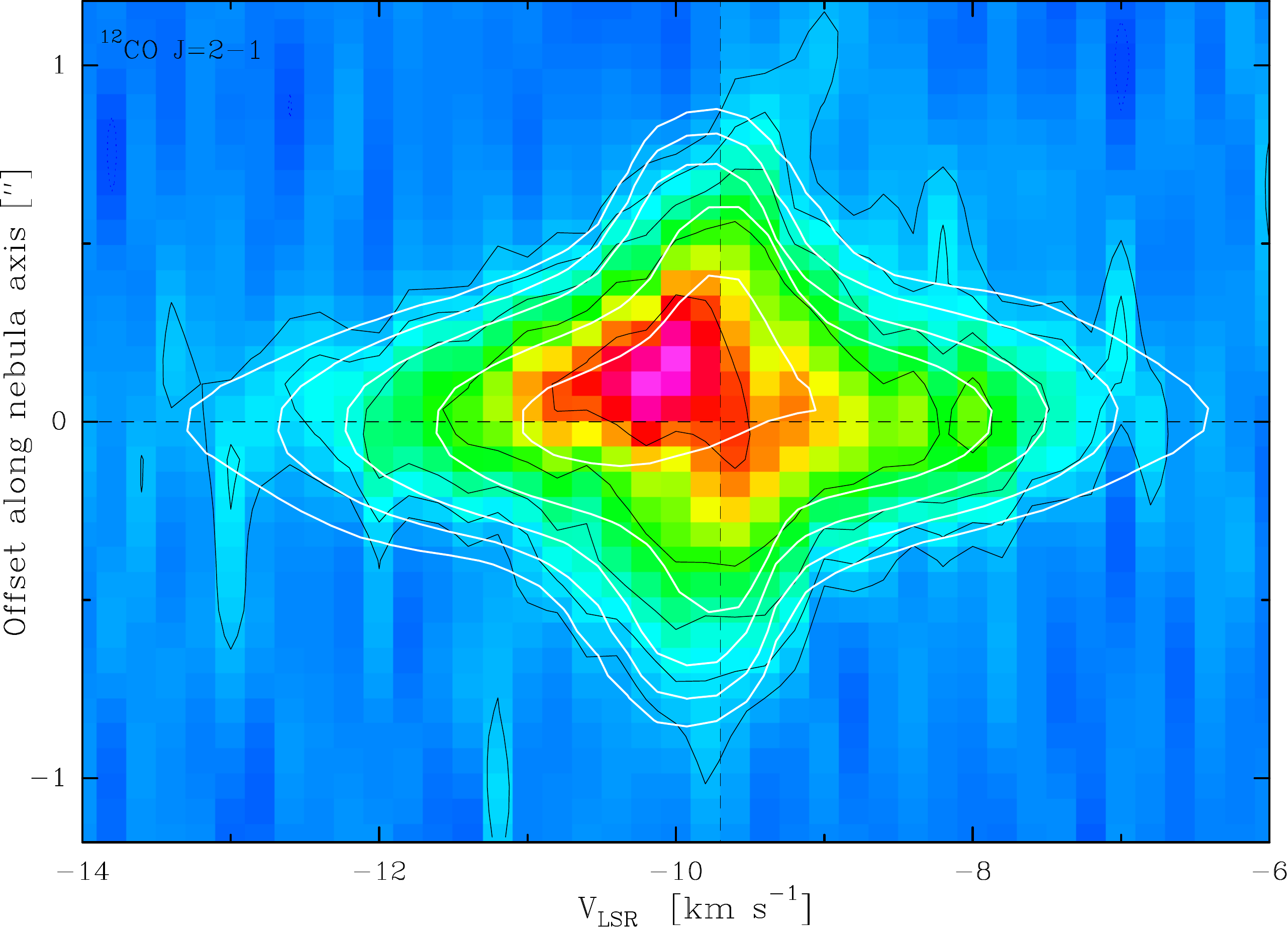}
        \end{minipage}
        \caption{\textit{Left:} PV diagram from our NOEMA maps of $^{12}$CO $J=2-1$ in~AC\,Her along the~equatorial direction of the~disk ($PA=136.1\degree$).
The contours are the~same as in~Figure\,\ref{fig:ac12mapas}.             
The dashed lines reveal the~systemic velocity and the~central position of the~source. 
Additionally, we~show the~synthetic PV diagram from our best-fit model
in white contours, to~be compared with observational data; the~scales and contours are the~same.
\textit{Right}: Same as in~\textit{left} but along the~axis direction of the~nebula ($PA=46.1\degree$).}
        \label{fig:acher12pv}
\end{figure}

\begin{figure}[H]
\includegraphics[scale=0.49]{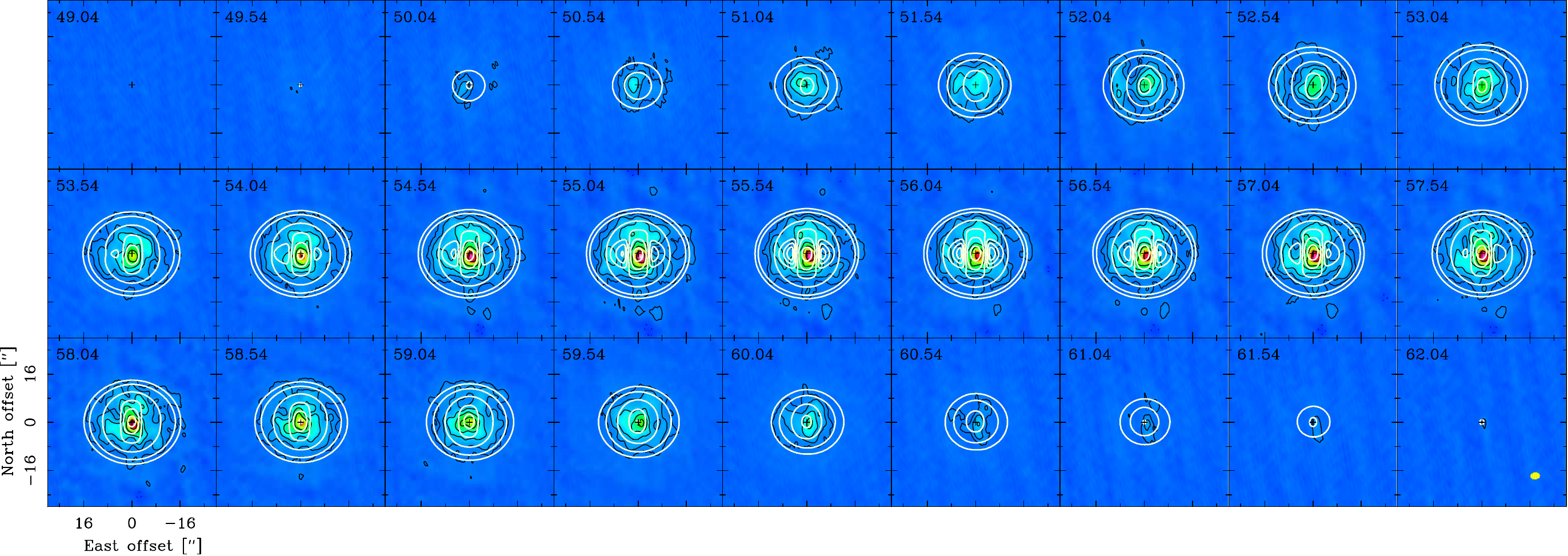}
 
\caption{Maps per velocity channel of R\,Sct in~$^{12}$CO $J=2-1$ emission. The~contours are $\pm$\,50, 100, 200, 400, 800, and 1600\,mJy\,beam$^{-1}$ with a~maximum emission of 2.4\,Jy\,beam$^{-1}$. The~LSR velocities are indicated in~each panel (upper-left corner) and the~beam size, 3.''12 \x2.''19, is shown in~the~last panel at the~bottom right corner (yellow ellipse).
We also show the~synthetic maps from our best-fit model in~white contours, to~be compared with observational data; the~scales and contours are the~same.}
    \label{fig:rs12mapas}  
\end{figure}

\begin{figure}[H]
        \begin{minipage}[b]{0.48\linewidth}
                \includegraphics[scale=0.25]{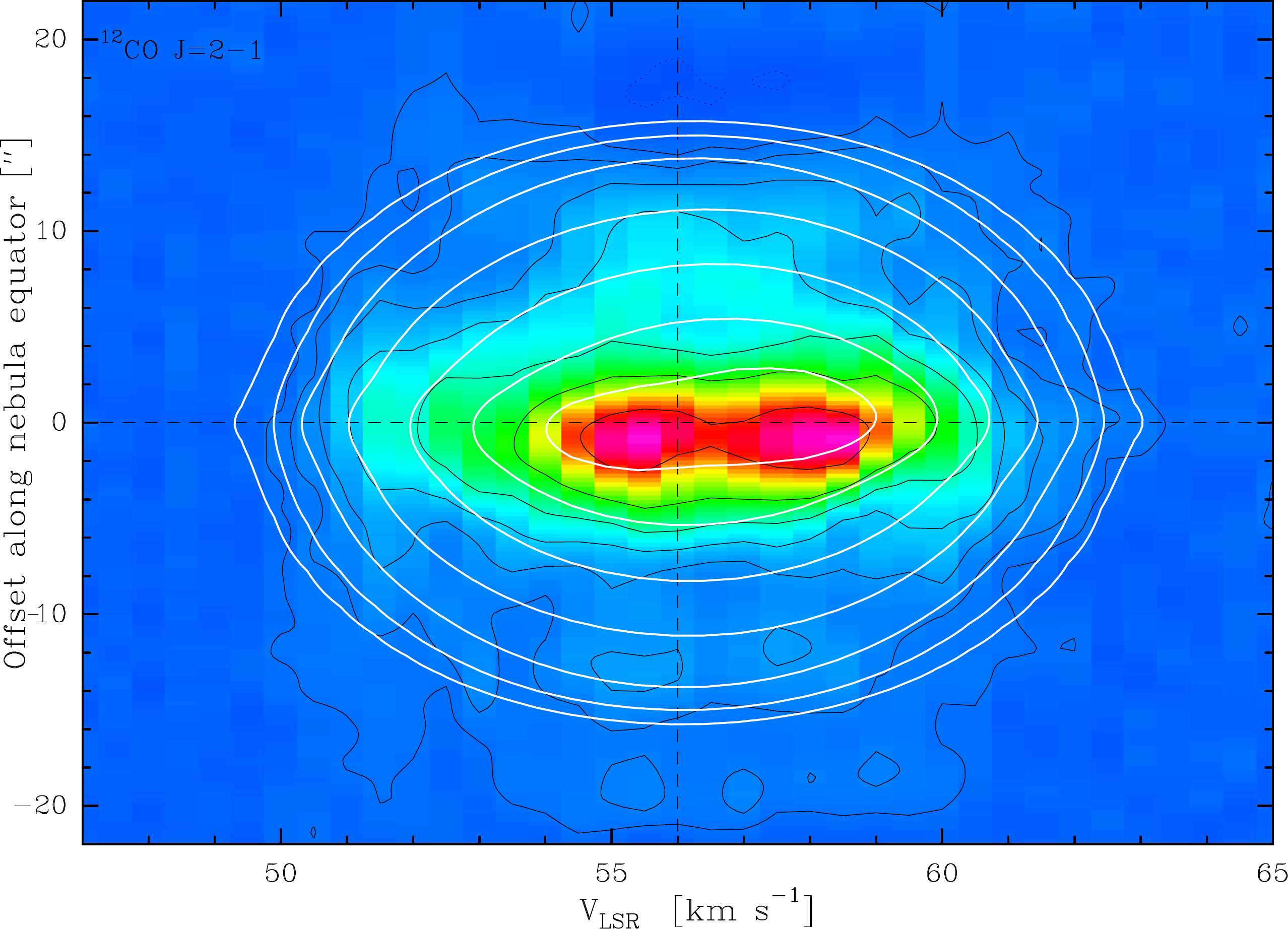}
        \end{minipage}
        \quad
        \begin{minipage}[b]{0.48\linewidth}
                \includegraphics[scale=0.25]{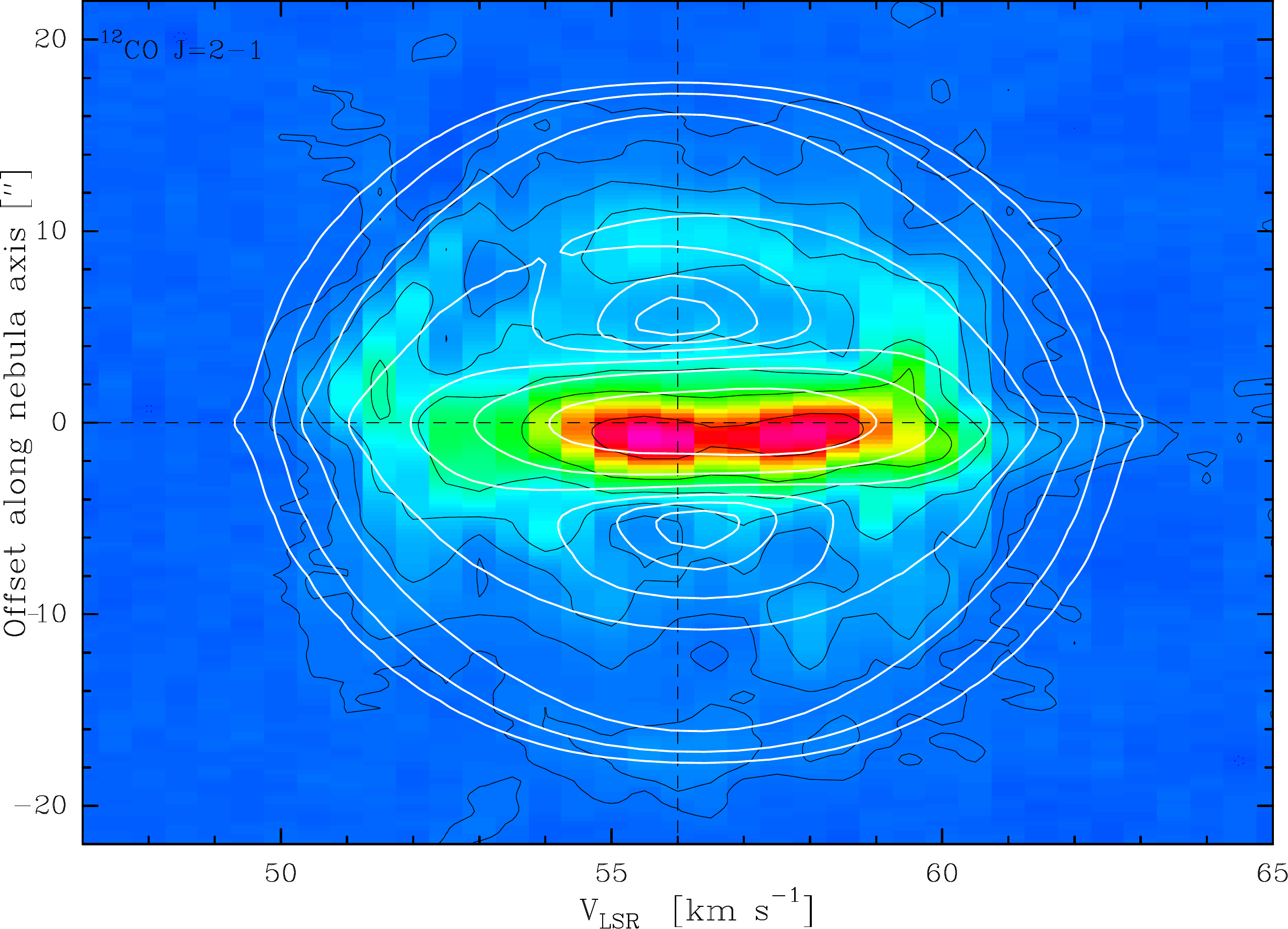}
        \end{minipage}
        \caption{\textit{Left:} PV diagram from our maps of $^{12}$CO $J=2-1$ in~R\,Sct along the~equatorial direction ($PA=0\degree$).
The contours are the~same as in~Figure\,\ref{fig:rs12mapas}.             
The dashed lines reveal the~systemic velocity and the~central position of the~source. 
Additionally, we~show the~synthetic PV diagram from our best-fit model
in white contours, to~be compared with observational data; the~scales and contours are the~same.
\textit{Right}: Same as in~\textit{left} but along the~axis direction ($PA=90\degree$).}
        \label{fig:rsct12pv}
\end{figure}

\begin{figure}[H]
        \includegraphics[scale=0.55]{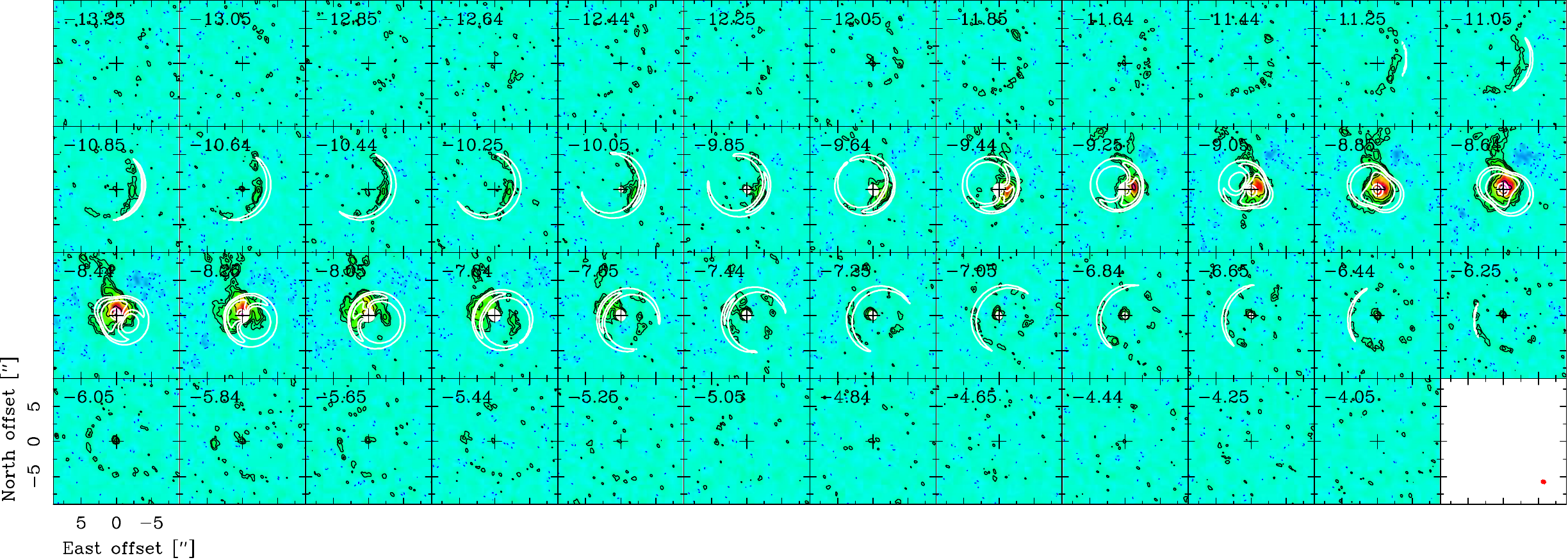}
   \caption{NOEMA maps per velocity channel of 89\,Her in~$^{13}$CO $J=2-1$ emission. The~contours are $\pm$\,11, 22, 44, 88, and~144\,mJy\,beam$^{-1}$ with a~maximum emission of 225\,mJy\,beam$^{-1}$. The~LSR velocities are indicated in~each panel (upper-left corner) and the~beam size, 0.''74\x 0.''56, is shown in~the~last panel at the~bottom right corner (red ellipse).
   We also show the~synthetic maps from our best-fit model in~white contours, to~be compared with observational data; the~scales and contours are the~same.}
    \label{fig:89hermapas}  
\end{figure}

\begin{figure}[H]
        \begin{minipage}[b]{0.48\linewidth}
                \includegraphics[scale=0.25]{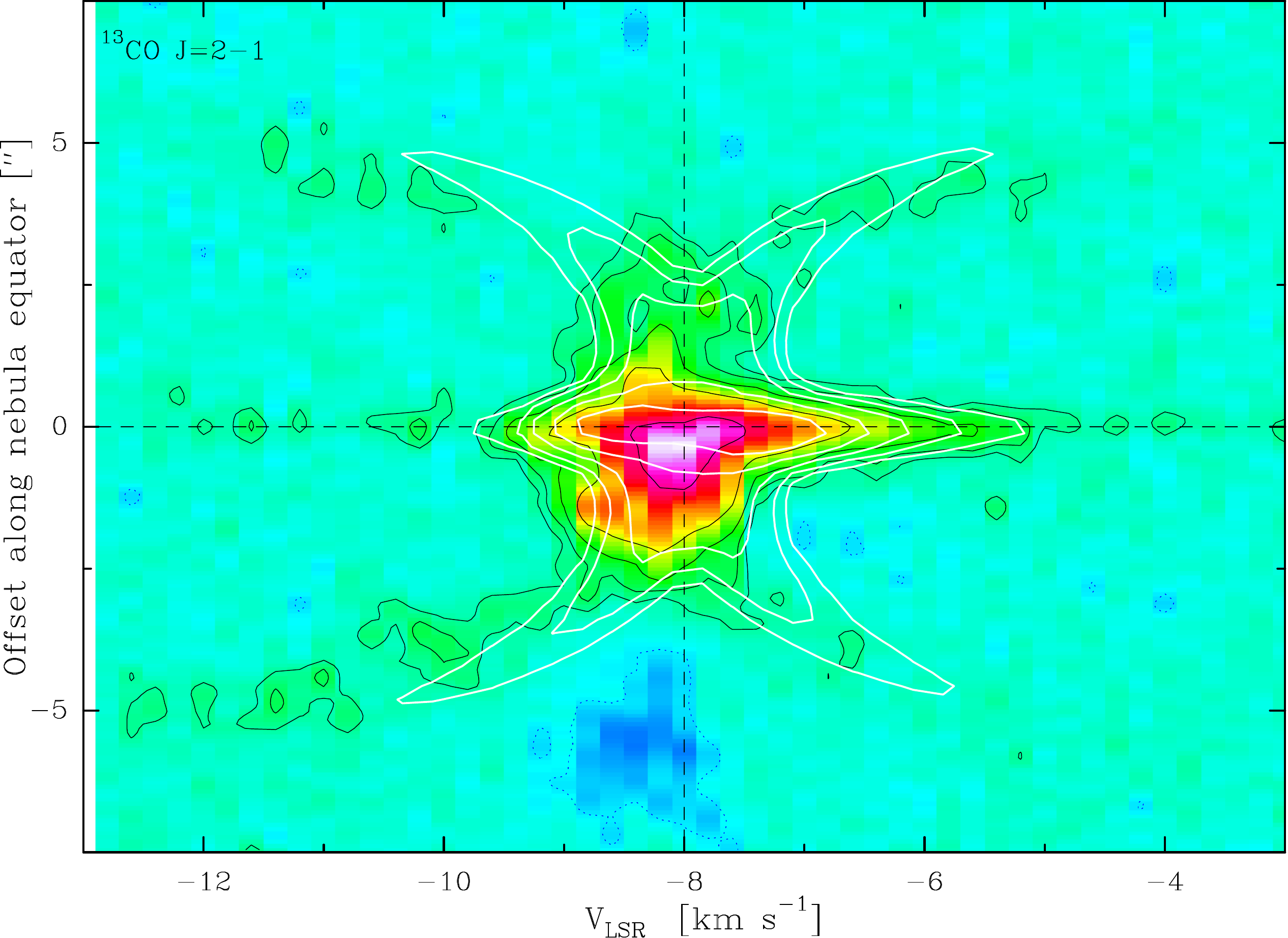}
        \end{minipage}
        \quad
        \begin{minipage}[b]{0.48\linewidth}
                \includegraphics[scale=0.25]{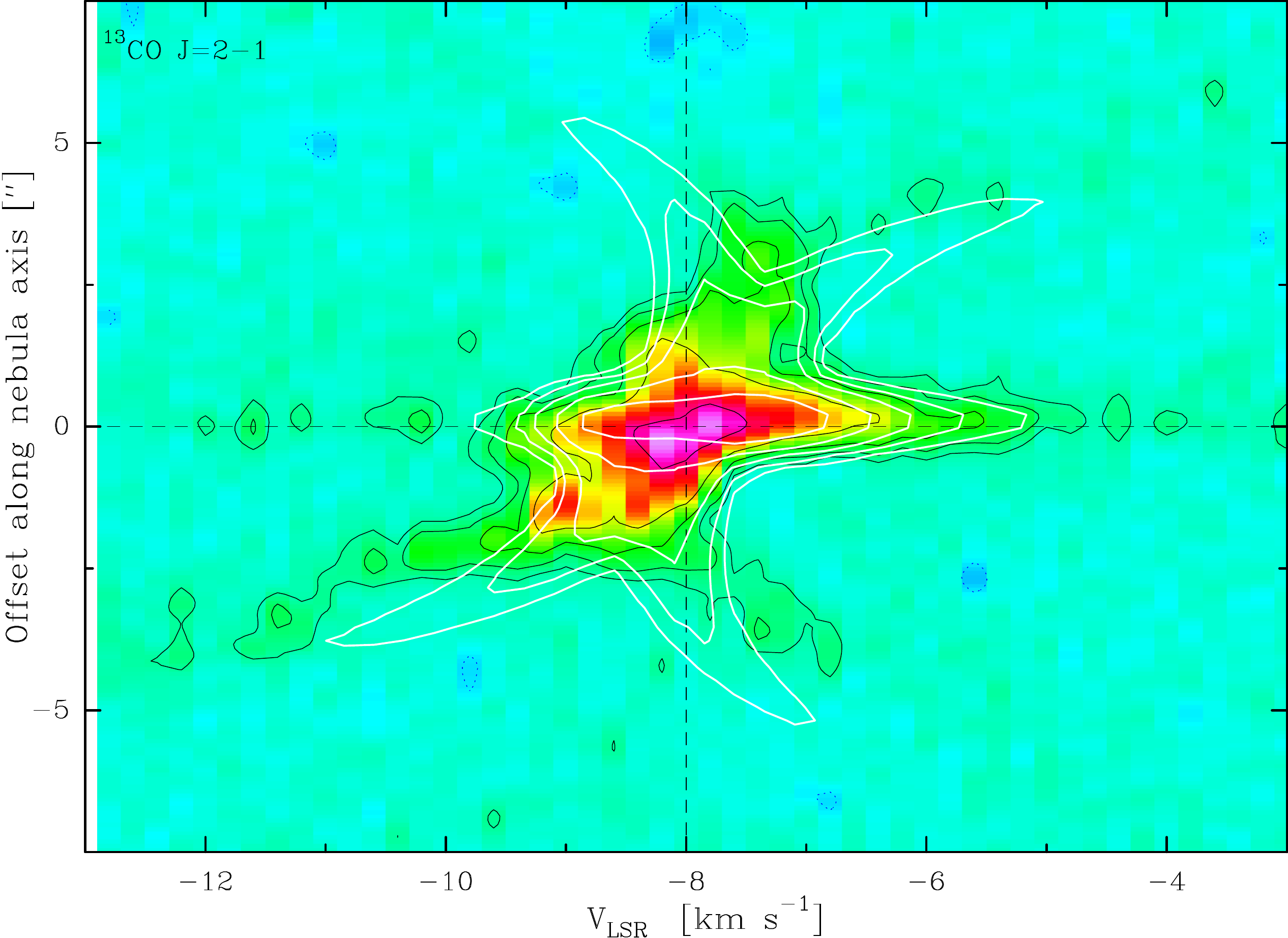}
        \end{minipage}
        \caption{ \textit{Left:} PV diagram from our NOEMA maps of $^{13}$CO $J=2-1$ in~89\,Her along the~equatorial direction of the~disk ($PA=150\degree$).
The contours are the~same as in~Figure\,\ref{fig:89hermapas}.             
The dashed lines reveal the~systemic velocity and the~central position of the~source. 
Additionally, we~show the~synthetic PV diagram from our best-fit model
in white contours, to~be compared with observational data; the~scales and contours are the~same.
\textit{Right}: Same as in~\textit{left} but along the~axis direction of the~nebula ($PA=60\degree$).}
        \label{fig:89her13pv}
\end{figure}

In this section, we~present the~results directly obtained from the~observations for AC\,Her, R\,Sct, and~89\,Her \citep[][]{gallardocava2021}. 
We show our NOEMA maps per velocity channel \linebreak  and position--velocity (PV) diagrams along~the equatorial rotating disk and along the~axis of the~nebula 
(see~Figures~\ref{fig:ac12mapas}, \ref{fig:acher12pv}, \ref{fig:rs12mapas}, \ref{fig:rsct12pv}, \ref{fig:89hermapas}, and ~\ref{fig:89her13pv}).


\unskip 

\unskip

\unskip

\unskip

\unskip

\subsection{AC\,Herculis}

The $^{12}$CO $J=2-1$ mm-wave interferometric maps are presented in~Figure\,\ref{fig:ac12mapas}.      
We can see in~the~left panel of Figure\,\ref{fig:acher12pv} the~PV diagram along the~equatorial direction, and it~very clearly shows the~characteristic signature of rotation with Keplerian dynamics. On~the contrary, the~analysis of the~PV diagram along the~nebula axis direction helps us to detect the~presence of an~axially outflowing component (see Figure\,\ref{fig:acher12pv} \textit{right}). 
A theoretical PV diagram along the~axis direction in~the~presence of a~disk with Keplerian dynamics might show emission with a~form close to a~rhombus, with~equal or very similar emission in~all~four quadrants of the~PV diagram.
Nevertheless, we~do not see equal emission in~the~four quadrants: we~see slightly inclined emission at central velocities at around $\pm$1$''$. This~fact can be explained by the~existence of an~expanding component that surrounds the~rotating~disk.

\subsection{R\,Scuti}

We present combined NOEMA maps and 30\,m maps of R\,Sct in~$^{12}$CO $J=2-1$ emission in~Figure\,\ref{fig:rs12mapas} and PV diagrams in~Figure\,\ref{fig:rsct12pv}. 
In both figures, we~clearly see two components: an~intense inner region and an~extended component of around $\sim$40$''$ surrounding the~inner region.
This extended and expanding component contains most of the~total nebular mass (see Section~\ref{secrsctmodel}).
The PV diagram along the~equatorial direction shows an~intense central clump in~the~innermost region of the~nebula that may represent the~unresolved rotating disk (see Figure\,\ref{fig:rsct12pv} \textit{left}). 
The velocity dispersion from the~inner (and unresolved) central condensation is similar to other post-AGB nebulae with disks, including a~significant lack of blueshifted emission (see, e.g.,~\citep{bujarrabal2016}).
The PV diagram along the~nebula axis reveals the~structure of the~nebula (see Figure\,\ref{fig:rsct12pv} \textit{right}): we~clearly see two large cavities at approximately $\pm$10$''$.
We see this type of structure in~other pPNe, such as M\,2$-$56 \citep{castrocarrizo2002} or M\,1$-$92 \citep{alcolea2007}.

\subsection{89\,Her}

We present NOEMA maps and PV diagrams of 89\,Her of $^{13}$CO $J=2-1$ emission (and $^{12}$CO $J=2-1$; see~\citep{gallardocava2021} for further details) in~Figures~\ref{fig:89hermapas} and \ref{fig:89her13pv}.
We see an~intense central clump and an~extended hourglass-shaped structure surrounding this central clump.
For a~distance of 1\,kpc, the~size of the~hourglass-like structure is, at~least, 10,000\,AU.

\subsection{Models} \label{models}

Our models consist of a~disk with present Keplerian dynamics and an~extended and expanding component  escaping from the~rotating disk and surrounding it.
The outflowing component can present different shapes, such as an~hourglass, an~ellipsoid, etc.
We assume LTE populations, which is a~reasonable assumption for low-$J$ rotational levels of CO transitions.
We consider potential laws for the~density ($n$) and rotational temperature ($T$). Additionally, we~also consider Keplerian dynamics in~the~rotating disk ($V_{rot\,K}$) and radial expansion in~the~extended component ($V_{exp}$).
We must highlight that our code produces results that can be quantitatively compared to~observations.

\begin{equation}
  n=n_{0}\left(\frac{r_{0}}{r}\right)^{\kappa_{n}},
\end{equation}

\begin{equation}
  T=T_{0}\left(\frac{r_{0}}{r}\right)^{\kappa_{T}},
\end{equation}

\begin{equation}
	V_{rot_{K}} = V_{rot_{K_{0}}}\sqrt{\frac{10^{16}}{r}},
\end{equation}

\begin{equation}
	V_{exp} = V_{exp_{0}} \frac{r}{10^{16}}.
\end{equation}

\subsubsection{AC\,Her}

Our proposed model for the~structure of AC\,Her (Figure\,\ref{fig:densidades}; see also  Figures~\ref{fig:ac12mapas}~and~\ref{fig:acher12pv}) is very similar to the~one found for the~Red\,Rectangle, IRAS\,08544$-$4431, and~IW\,Car (see~\citep{bujarrabal2016,bujarrabal2017, bujarrabal2018}).
The total mass of the~nebula is 8.3\x 10$^{-4}$\,M$_{\odot}$.
Our model predicts that the~mass of the~outflow must be $\leq$5\% of the~total mass.
Thus, AC\,Her is clearly a~binary post-AGB star surrounded by a~disk-dominated nebula, due to the fact that the~mass of the~Keplerian disk is, at~least, 19 times larger than that of the~outflow.
The Keplerian rotation velocity field of the~disk is compatible with a~central total stellar mass of $\sim$1\,M$_{\odot}$.

\subsubsection{R\,Sct} \label{secrsctmodel}

\begin{figure}[H]
    	\includegraphics[scale=0.42]{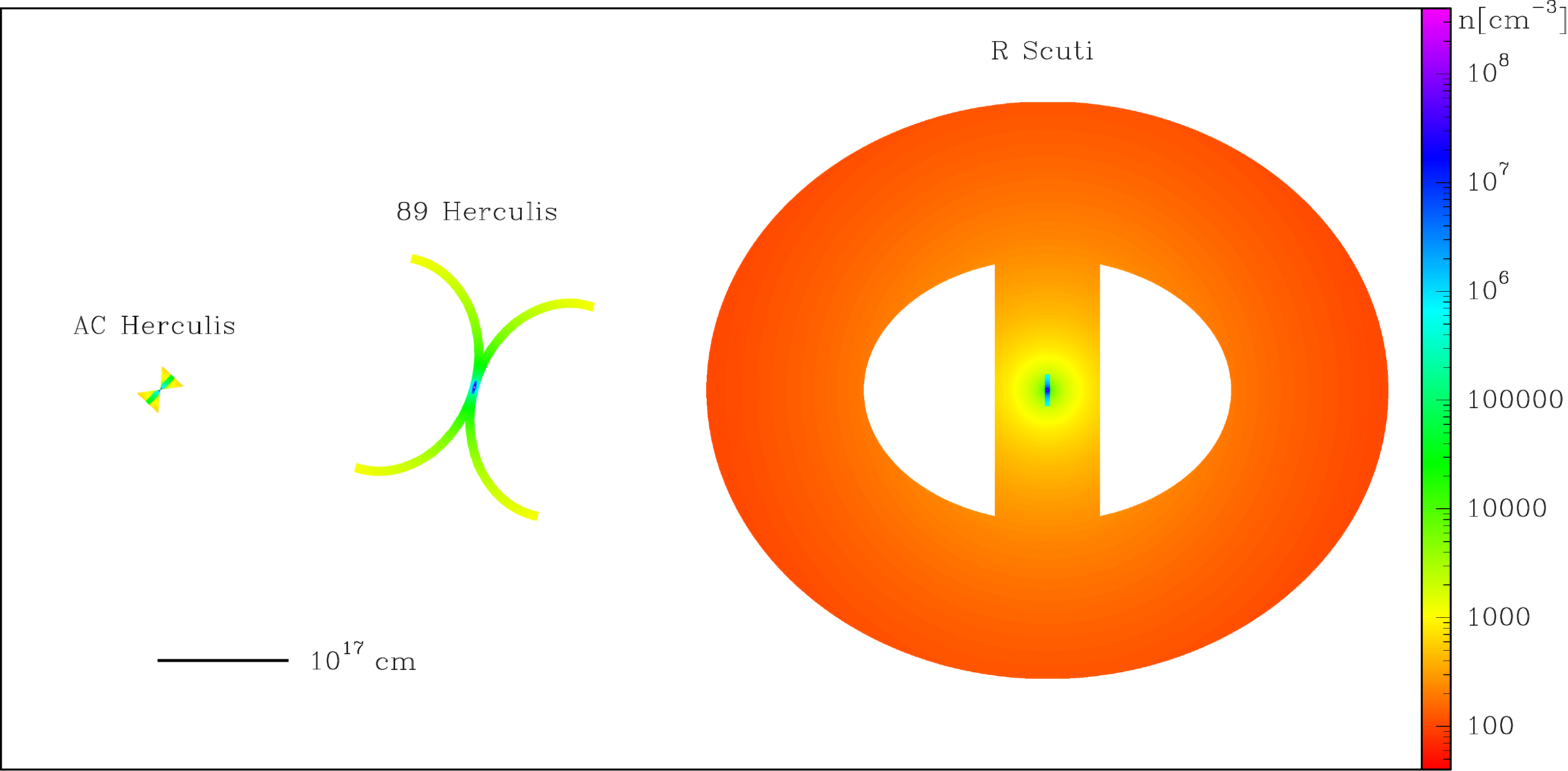}
    \caption{Density distribution and structure of our best-fit model for the~disk and outflow of AC\,Her, 89\,her, and~R\,Sct as seen from the~observer. All~models are shown with the~same scale so that we~can see their relative~sizes.}
        \label{fig:densidades}
\end{figure}

Our proposed model for the~structure of R\,Sct is presented (Figure\,\ref{fig:densidades}; see also  Figures~\ref{fig:rs12mapas}~and~\ref{fig:rsct12pv}). There are slight differences between the PV diagrams, but~all of them are accounted for in~our uncertainties. We~note that our best-fit model cannot be very different from other models.
The nature of R\,Sct is not yet clear, but~our interferometric maps firmly suggest that this source is also a~binary post-AGB star surrounded by a~disk with Keplerian dynamics and by a~high-mass extended and expanding component.
The mass of the~nebula is found to be $\sim$\,3.2\x10$^{-2}$\,M$_{\odot}$ and approximately $\sim$25\% of the~nebular material would be placed in~the~rotating disk.
This fact, together with the~large size of the~outflow, allows us to classify this source as an~outflow-dominated post-AGB nebula.
The disk with Keplerian dynamics is compatible with a~central stellar mass of 1.7\,M$_{\odot}$.

\unskip    

\subsubsection{89\,Her}

The total mass of the~nebula around 89\,Her is 1.4\x 10$^{-2}$\,M$_{\odot}$ and our proposed model predicts that the~mass of the~hourglass must be $\sim$50\% of the~total mass (Figure\,\ref{fig:densidades}; see also  Figures~\ref{fig:89hermapas}~and~\ref{fig:89her13pv}).
Thus, this source is in~between the~disk- and outflow-dominated sources.
We find that the~disk with Keplerian dynamics is compatible with a~central stellar mass of 1.7\,M$_{\odot}$.

\section{First Molecular Survey in~Binary Post-AGB~Stars} \label{mole}
The chemistry of this kind of binary post-AGB source with rotating disks is practically unknown. We~present a~very deep and wide survey of radio lines in~ten of our sources: AC\,Her, the~Red\,Rectangle, 89\,Her, HD\,52961, IRAS\,19157$-$0257\, IRAS\,18123+0511, IRAS\,19125+0343, AI\,CMi, IRAS\,20056+1834, and~R\,Sct. All~of them have been observed at 7 and 13 mm, and~most of them have been also observed in~the~1.3, 2, and~3\,mm bands; see Table\,\ref{lineas} (see~\citep{gallardocava2022}).

\begin{table}[H]
\caption{Molecular transitions detected in~this~work.}\label{lineas}
\tablesize{\footnotesize}
\newcolumntype{C}{>{\centering\arraybackslash}X}
\begin{tabularx}{\textwidth}{>{\centering}m{0.8cm}C>{\centering}m{2.5cm}CCCCC}
\toprule
\multicolumn{4}{c}{\textbf{O-Bearing Molecules}} & \multicolumn{4}{c}{\textbf{C-Bearing Molecules}}\\\cmidrule{1-8}
{\textbf{Species}} & \multicolumn{2}{c}{\textbf{Transition}} & \boldmath{$\nu$} \textbf{[MHz]}& {\textbf{Molecule}} & \multicolumn{2}{c}{\textbf{Transition}} & \boldmath{$\nu$} \textbf{[MHz]}\\
\midrule

$^{28}$SiO & $v=0$ & $J=1-0$ & 43,423.85  & HCN & $v=0$ & $J=1-0$ & 88,630.42 \\
 &  $v=0$ & $J=2-1$ & 86,846.99  & CS & $v=0$ & $J=3-2$ & 146,969.00  \\
 &  $v=0$ & $J=5-4$ & 217,104.98 & SiS & $v=0$ & $J=5-4$ & 90,771.56  \\
 &  $v=1$ & $J=1-0$ & 43,122.08 \\
 &  $v=1$ & $J=2-1$ & 86,243.37   \\ 
 &  $v=2$ & $J=1-0$ & 42,820.59  \\

SO & $v=0$ & $J_{N}=6_{5}-5_{4}$ & 219,949.44  \\

H$_{2}$O & $v=0$ & $J_{Ka,\,Kc}=6_{1,\,6}-5_{2,\,3}$ & 22,235.08 \\

\bottomrule
\end{tabularx}
\end{table}
\unskip


\subsection{Molecular~Richness}

We show in~Figure~\ref{fig:moleculas_raras} integrated intensity ratios between the~main rare molecules (SO,~SiO, SiS, CS, and~HCN) and CO ($^{13}$CO $J=2-1$ and $^{12}$CO $J=1-0$). Additionally, we~also compare these molecular integrated intensities with the~12, 25, and~60\,$\upmu$m IR emission. The~averaged of our results are represented with black horizontal lines.
We compare our results with the~molecular emission of AGB stars (blue and red horizontal lines represent averaged values of the~molecular emission for O- and C-rich AGB stars, respectively). We~note that the~average of our results always presents low molecular emission in~molecules other than CO. Note the~large range (logarithmic scale) of intensity ratios. These~low intensities are more remarkable in~the~disk-dominated sources, such as AC\,Her and the~Red\,Rectangle (see~\citep{gallardocava2021, gallardocava2022}).

\begin{figure}[H]
\includegraphics[scale=0.7]{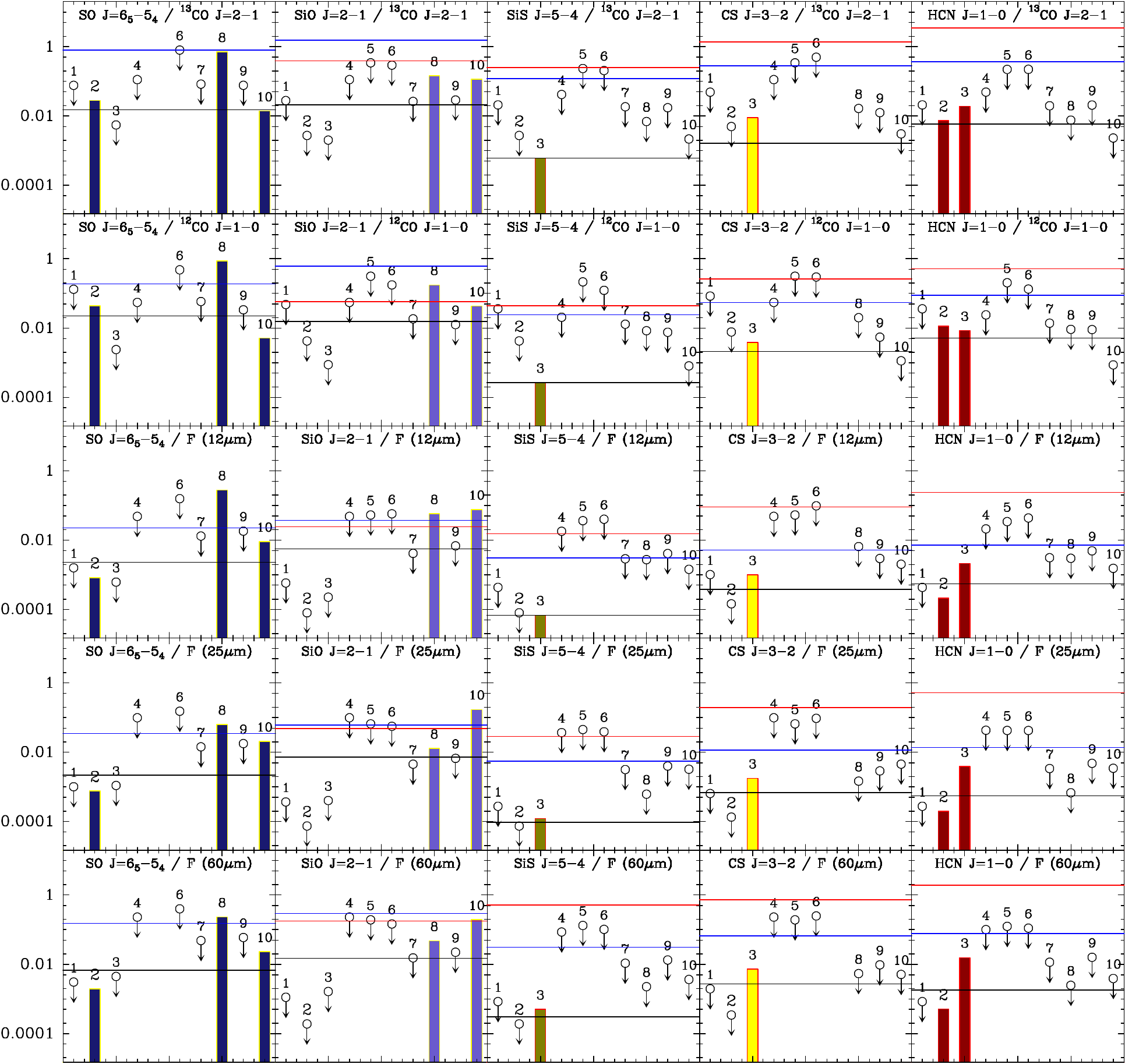}
\caption{Ratios of integrated intensities of molecules (SO, SiO, SiS, CS, and~HCN) and IR emission (12, 25, and~60\,$\upmu$m) in~our sources. The~binary post-AGB stars are ordered by increasing outflow dominance: 1---AC\,Her, 2---Red\,Rectangle, 3---89\,Herculis, 4---HD\,52961, 5---IRAS\,19157$-$0257, \mbox{6---IRAS\,18123+0511}, 7---IRAS\,19125+0343, 8---AI\,CMi, 9---IRAS\,20056+1834, and~10---R\,Sct.
Empty circles with arrows represent upper limits. Our results are averaged (black lines) and are compared with averaged values for O- and C-rich AGB CSEs (blue and red lines, values taken from~\citep{bujarrabal1994a, bujarrabal1994b}). Our sample of nebulae around binary post-AGB stars clearly presents low molecular emission in~molecules other than CO.}
 \end{figure}

\subsection{The Discrimination between O- and C-Rich~Envelopes}

Evolved stars present a~O/C\,>\,1 and O/C\,<\,1 chemistry, so they can be classified as O- and C-rich environments, respectively. 
The O/C abundance ratio has important effects on~the~molecular abundances. 
The lines of O-bearing molecules are much more intense in~O-rich environments than in~the~C-rich ones (such as SiO and SO). On~the contrary, the~lines of C-bearing molecules are much more intense in~C-rich environments \mbox{(see,~e.g.,~\citep{bujarrabal1994a, bujarrabal1994b})}. Additionally, SiO and H$_{2}$O maser emission is exclusive to O-rich environments (see,~e.g.,~\citep{kim2019}).

\begin{figure}[H]
\includegraphics[scale=0.6]{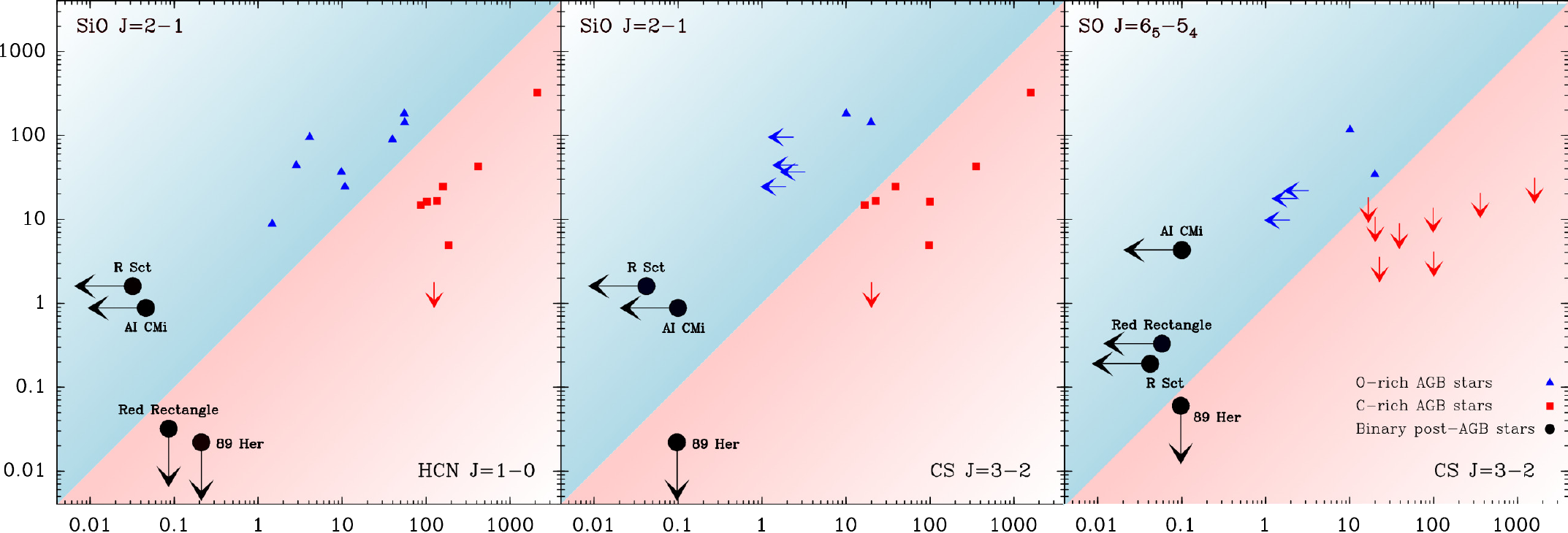}
\caption{Integrated intensities of pairs of molecular transitions in~binary post-AGB stars (black circles, upper limits are represented with arrows), as~well in~O- and C-rich AGB CSEs (blue and red squares). 
O- and C-rich environment areas are represented in~blue and red, respectively.
The integrated intensities of the~transitions are expressed in~Jy\,km\,s$^{-1}$ and in~logarithmic scale.}
    \label{fig:cuadros_mol_mol}  
\end{figure}

\textls[-15]{We analyze the integrated intensities of pairs of molecular transitions, and~this analysis is crucial to distinguish between O- and C-rich environments (see Figure~\ref{fig:cuadros_mol_mol}). 
When an~O-bearing molecule is compared with a~C-bearing one, we~find that the integrated intensities are larger in~O-rich than in~C-rich sources. Our results are compared with CSEs around AGB stars, because~they are prototypical environments rich in~molecules.}

Based on~the~maser detection of O-bearing molecules (SiO maser emission in~R\,Sct, AI\,CMi, and~IRAS\,20056+1834; H$_{2}$O maser emission in~R\,Sct and AI\,CMi), we~classify some of our sources as O-rich.
On the~contrary, and~based on~the~integrated intensity ratios, we~classify the~nebula around 89\,Her as C-rich (see Figure\,\ref{fig:cuadros_mol_mol}).
Therefore, the~nebula around AC\,Her, the~Red\,Rectangle, AI\,CMi, IRAS\,20056+1834, and~R\,Sct presents a~O/C\,>\,1 chemistry, while 89\,Her presents a~O/C\,<\,1 environment \citep{gallardocava2022}.

\section{Conclusions}\label{sec5}

There is a~class of post-AGB star that is part of a~binary system with a~significant NIR excess that is surrounded by a~disk with Keplerian dynamics and an~extended and expanding component composed of gas escaping from the~disk and surrounding~it.

Based on our observational data and model results, we~find disk-dominated sources that present $\geq$85\% of the~total nebular mass located in~the~Keplerian disk. This~is the~case of AC\,Herculis.
We also find a~subclass of these binary post-AGB stars, in~which the~disk contains $\sim$25\% of the~total mass of the~nebula, such as R\,Scuti.
The extended components of these outflow-dominated sources are mainly composed of cold gas.
Moreover, our NOEMA maps and modeling suggest that the~nebula around 89\,Her is in~an intermediate case~between both the~disk- and the~outflow-dominated sources, since around 50\% of the~nebular mass is located in~the~rotating disk. See Section~\ref{obs} for further details. 
HD\,52961 and IRAS\,1957$-$0247 would also belong to this intermediate case.
However, the~existence of this intermediate type is not clear, because~these objects were classified as intermediate sources under high uncertainties and  they could belong to either subclass: the~disk- or the~outflow-dominated sources.
In the~case of 89\,Her, our new 30\,m\,IRAM on-the-fly observations recover all the~filtered flux. These~maps show a~larger hourglass-like structure compared to that in~the NOEMA maps. According to these new maps and preliminary results, the~hourglass-like structure around 89\,Her could contain most of the~material (Gallardo Cava~et~al., in~prep). 

We present the~first survey in~the~search for molecules other than CO in~binary post-AGB stars surrounded by Keplerian disks (see Section~\ref{mole}).
The emission of molecules other than CO in~our sources is low and this fact is especially remarkable in~the~disk-dominated nebulae.
Additionally, and according to our analysis, we~catalog the~chemistry of 89\,Her as C-rich. On~the contrary, we~find O-rich environments in~AC\,Her, the~Red\,Rectangle, AI\,CMi, IRAS\,20056+1834, and~R\,Sct.

\newpage



\vspace{6pt}



\authorcontributions{
Conceptualization, I.G.C., V.B., and J.A.; methodology, I.G.C, V.B., and J.A.; software, I.G.C. and V.B.; validation, I.G.C., V.B., J.A., M.G.-G., A.C.-C., H.V.W., and M.S.-G.;  formal analysisI.G.C., V.B., J.A., and H.V.W.; investigation, I.G.C., V.B., and J.A.; resources, I.G.C., V.B., J.A., M.G.-G., A.C.-C., H.V.W., and M.S.-G.; data curation, I.G.C., V.B., J.A., M.G.-G., and A.C.-C.; writing—original draft preparation, I.G.C.; writing—review and editing, I.G.C., V.B., and J.A.; visualization, I.G.C., V.B., and J.A., supervision, I.G.C., V.B., and J.A.; project administration, J.A., funding acquisition, J.A. and V.B.
All authors have agreed to the published version of the manuscript.}
 

\funding{This work is part of the~AxiN and EVENTs\,/\,NEBULAE\,WEB research programs supported by Spanish AEI grants AYA\,2016-78994-P and PID2019-105203GB-C21.
I.G.C. acknowledges Spanish MICIN for the~funding support of~BES2017-080616.
}

\dataavailability{Not applicable.} 

\conflictsofinterest{The authors declare no conflicts of~interest.}



%



\begin{adjustwidth}{-\extralength}{0cm}
\reftitle{References}



\begin{thebibliography}{999}

\bibitem[{Gallardo Cava} \em{et~al.}(2021){Gallardo Cava}, {G{\'o}mez-Garrido},
  {Bujarrabal}, {Castro-Carrizo}, {Alcolea}, and~{Van
  Winckel}]{gallardocava2021}
{Gallardo Cava}, I.; {G{\'o}mez-Garrido}, M.; {Bujarrabal}, V.;
  {Castro-Carrizo}, A.; {Alcolea}, J.; {Van Winckel}, H.
\newblock {Keplerian disks and outflows in~post-AGB stars: AC Herculis, 89
  Herculis, IRAS 19125+0343, and~R Scuti}.
\newblock {\em Astron. Astrophys.} {\bf 2021}, {\em 648},~A93.
\newblock {{https://doi.org/10.1051/0004-6361/202039604}}.

\bibitem[{Gallardo Cava} \em{et~al.}(2022){Gallardo Cava}, {Bujarrabal},
  {Alcolea}, {G{\'o}mez-Garrido}, and
  {Santander-Garc{\'\i}a}]{gallardocava2022}
{Gallardo Cava}, I.; {Bujarrabal}, V.; {Alcolea}, J.; {G{\'o}mez-Garrido}, M.;
  {Santander-Garc{\'\i}a}, M.
\newblock {Chemistry of nebulae around binary post-AGB stars: A molecular
  survey of mm-wave lines}.
\newblock {\em Astron. Astrophys.} {\bf 2022}, {\em 659},~A134.
\newblock {{https://doi.org/10.1051/0004-6361/202142339}}.

\bibitem[{Van Winckel}(2003)]{vanwinckel2003}
\textls[-30]{{Van Winckel}, H.
\newblock {Post-AGB Stars}.
\newblock {\em Annu. Rev. Astron. Astrophys.} {\bf 2003}, {\em 41},~391--427.
\newblock {{https://doi.org/10.1146/annurev.astro.41.071601.170018}}.}

\bibitem[{Oomen} \em{et~al.}(2018){Oomen}, {Van Winckel}, {Pols}, {Nelemans},
  {Escorza}, {Manick}, {Kamath}, and~{Waelkens}]{oomen2018}
{Oomen}, G.M.; {Van Winckel}, H.; {Pols}, O.; {Nelemans}, G.; {Escorza}, A.;
  {Manick}, R.; {Kamath}, D.; {Waelkens}, C.
\newblock {Orbital properties of binary post-AGB stars}.
\newblock {\em Astron. Astrophys.} {\bf 2018}, {\em 620},~A85.
\newblock {{https://doi.org/10.1051/0004-6361/201833816}}.

\bibitem[{Gielen} \em{et~al.}(2011){Gielen}, {Bouwman}, {van Winckel}, {Lloyd
  Evans}, {Woods}, {Kemper}, {Marengo}, {Meixner}, {Sloan}, and
  {Tielens}]{gielen2011a}
{Gielen}, C.; {Bouwman}, J.; {van Winckel}, H.; {Lloyd Evans}, T.; {Woods},
  P.M.; {Kemper}, F.; {Marengo}, M.; {Meixner}, M.; {Sloan}, G.C.; {Tielens},
  A.G.G.M.
\newblock {Silicate features in~Galactic and extragalactic post-AGB discs}.
\newblock {\em Astron. Astrophys.} {\bf 2011}, {\em 533},~A99.
\newblock {{https://doi.org/10.1051/0004-6361/201117364}}.

\bibitem[{Jura}(2003)]{jura2003}
{Jura}, M.
\newblock {A Flared, Orbiting, Dusty Disk around HD 233517}.
\newblock {\em Astrophys. J.} {\bf 2003}, {\em 582},~1032--1035.
\newblock {{https://doi.org/10.1086/344704}}.

\bibitem[{Sahai} \em{et~al.}(2011){Sahai}, {Claussen}, {Schnee}, {Morris}, and
  {S{\'a}nchez Contreras}]{sahai2011}
{Sahai}, R.; {Claussen}, M.J.; {Schnee}, S.; {Morris}, M.R.; {S{\'a}nchez
  Contreras}, C.
\newblock {An Expanded Very Large Array and CARMA Study of Dusty Disks and
  Torii with Large Grains in~Dying Stars}.
\newblock {\em Astrophys. J.} {\bf 2011}, {\em 739},~L3.
\newblock {{https://doi.org/10.1088/2041-8205/739/1/L3}}.

\bibitem[{Hillen} \em{et~al.}(2017){Hillen}, {Van Winckel}, {Menu}, {Manick},
  {Debosscher}, {Min}, {de Wit}, {Verhoelst}, {Kamath}, and
  {Waters}]{hillen2017}
{Hillen}, M.; {Van Winckel}, H.; {Menu}, J.; {Manick}, R.; {Debosscher}, J.;
  {Min}, M.; {de Wit}, W.J.; {Verhoelst}, T.; {Kamath}, D.; {Waters}, L.B.F.M.
\newblock {A mid-IR interferometric survey with MIDI/VLTI: Resolving the
  second-generation protoplanetary disks around post-AGB binaries}.
\newblock {\em Astron. Astrophys.} {\bf 2017}, {\em 599},~A41.
\newblock {{https://doi.org/10.1051/0004-6361/201629161}}.

\bibitem[{Kluska} \em{et~al.}(2019){Kluska}, {Van Winckel}, {Hillen}, {Berger},
  {Kamath}, {Le Bouquin}, and~{Min}]{kluska2019}
{Kluska}, J.; {Van Winckel}, H.; {Hillen}, M.; {Berger}, J.P.; {Kamath}, D.;
  {Le Bouquin}, J.B.; {Min}, M.
\newblock {VLTI/PIONIER survey of disks around post-AGB binaries. Dust
  sublimation physics rules}.
\newblock {\em Astron. Astrophys.} {\bf 2019}, {\em 631},~A108.
\newblock {{https://doi.org/10.1051/0004-6361/201935785}}.

\bibitem[{Bujarrabal} \em{et~al.}(2013){Bujarrabal}, {Alcolea}, {Van Winckel},
  {Santander-Garc{\'\i}a}, and~{Castro-Carrizo}]{bujarrabal2013a}
{Bujarrabal}, V.; {Alcolea}, J.; {Van Winckel}, H.; {Santander-Garc{\'\i}a},
  M.; {Castro-Carrizo}, A.
\newblock {Extended rotating disks around post-AGB stars}.
\newblock {\em Astron. Astrophys.} {\bf 2013}, {\em 557},~A104.
\newblock {{https://doi.org/10.1051/0004-6361/201322015}}.

\bibitem[{Guilloteau} \em{et~al.}(2013){Guilloteau}, {Di Folco}, {Dutrey},
  {Simon}, {Grosso}, and~{Pi{\'e}tu}]{guilloteau2013}
{Guilloteau}, S.; {Di Folco}, E.; {Dutrey}, A.; {Simon}, M.; {Grosso}, N.;
  {Pi{\'e}tu}, V.
\newblock {A sensitive survey for $^{13}$CO, CN, H$_{2}$CO, and~SO in~the~disks
  of T Tauri and Herbig Ae stars}.
\newblock {\em Astron. Astrophys.} {\bf 2013}, {\em 549},~A92.
\newblock {{https://doi.org/10.1051/0004-6361/201220298}}.

\bibitem[{Bujarrabal} \em{et~al.}(2016){Bujarrabal}, {Castro-Carrizo},
  {Alcolea}, {Santand er-Garc{\'\i}a}, {van Winckel}, and~{S{\'a}nchez
  Contreras}]{bujarrabal2016}
{Bujarrabal}, V.; {Castro-Carrizo}, A.; {Alcolea}, J.; {Santand
  er-Garc{\'\i}a}, M.; {van Winckel}, H.; {S{\'a}nchez Contreras}, C.
\newblock {Further ALMA observations and detailed modeling of the~Red
  Rectangle}.
\newblock {\em Astron. Astrophys.} {\bf 2016}, {\em 593},~A92.
\newblock {{https://doi.org/10.1051/0004-6361/201628546}}.

\bibitem[{Castro-Carrizo} \em{et~al.}(2002){Castro-Carrizo}, {Bujarrabal},
  {S{\'a}nchez Contreras}, {Alcolea}, and~{Neri}]{castrocarrizo2002}
{Castro-Carrizo}, A.; {Bujarrabal}, V.; {S{\'a}nchez Contreras}, C.; {Alcolea},
  J.; {Neri}, R.
\newblock {The structure and dynamics of the~molecular envelope of M 2-56}.
\newblock {\em Astron. Astrophys.} {\bf 2002}, {\em 386},~633--645.
\newblock {{https://doi.org/10.1051/0004-6361:20020220}}.

\bibitem[{Alcolea} \em{et~al.}(2007){Alcolea}, {Neri}, and
  {Bujarrabal}]{alcolea2007}
{Alcolea}, J.; {Neri}, R.; {Bujarrabal}, V.
\newblock {Minkowski's footprint revisited. Planetary nebula formation from a
  single sudden event?}
\newblock {\em Astron. Astrophys.} {\bf 2007}, {\em 468},~L41--L44.
\newblock {{https://doi.org/10.1051/0004-6361:20066956}}.

\bibitem[{Bujarrabal} \em{et~al.}(2017){Bujarrabal}, {Castro-Carrizo},
  {Alcolea}, {Van Winckel}, {S{\'a}nchez Contreras}, and~{Santand
  er-Garc{\'\i}a}]{bujarrabal2017}
{Bujarrabal}, V.; {Castro-Carrizo}, A.; {Alcolea}, J.; {Van Winckel}, H.;
  {S{\'a}nchez Contreras}, C.; {Santand er-Garc{\'\i}a}, M.
\newblock {A second post-AGB nebula that contains gas in~rotation and in
  expansion: ALMA maps of IW Carinae}.
\newblock {\em Astron. Astrophys.} {\bf 2017}, {\em 597},~L5.
\newblock {{https://doi.org/10.1051/0004-6361/201629317}}.

\bibitem[{Bujarrabal} \em{et~al.}(2018){Bujarrabal}, {Castro-Carrizo}, {Van
  Winckel}, {Alcolea}, {S{\'a}nchez Contreras}, {Santander-Garc{\'\i}a}, and
  {Hillen}]{bujarrabal2018}
{Bujarrabal}, V.; {Castro-Carrizo}, A.; {Van Winckel}, H.; {Alcolea}, J.;
  {S{\'a}nchez Contreras}, C.; {Santander-Garc{\'\i}a}, M.; {Hillen}, M.
\newblock {High-resolution observations of IRAS 08544-4431. Detection of a~disk
  orbiting a~post-AGB star and of a~slow disk wind}.
\newblock {\em Astron. Astrophys.} {\bf 2018}, {\em 614},~A58.
\newblock {{https://doi.org/10.1051/0004-6361/201732422}}.

\bibitem[{Bujarrabal} \em{et~al.}(1994{\natexlab{a}}){Bujarrabal}, {Fuente},
  and {Omont}]{bujarrabal1994a}
{Bujarrabal}, V.; {Fuente}, A.; {Omont}, A.
\newblock {The Discrimination between O- and C-rich Circumstellar Envelopes
  from Molecular Observations}.
\newblock {\em Astrophys. J.} {\bf 1994}, {\em 421},~L47.
\newblock {{https://doi.org/10.1086/187184}}.

\bibitem[{Bujarrabal} \em{et~al.}(1994{\natexlab{b}}){Bujarrabal}, {Fuente},
  and {Omont}]{bujarrabal1994b}
{Bujarrabal}, V.; {Fuente}, A.; {Omont}, A.
\newblock {Molecular observations of O- and C-rich circumstellar envelopes}.
\newblock {\em Astron. Astrophys.} {\bf 1994}, {\em 285},~247--271.

\bibitem[{Kim} \em{et~al.}(2019){Kim}, {Cho}, {Bujarrabal}, {Imai}, {Dodson},
  {Yoon}, and~{Zhang}]{kim2019}
{Kim}, J.; {Cho}, S.H.; {Bujarrabal}, V.; {Imai}, H.; {Dodson}, R.; {Yoon},
  D.H.; {Zhang}, B.
\newblock {Time variations of H$_{2}$O and SiO masers in~the~proto-Planetary
  Nebula OH 231.8+4.2}.
\newblock {\em Mon. Not. R. Astron. Soc.} {\bf 2019}, {\em 488},~1427--1445.
\newblock {{https://doi.org/10.1093/mnras/stz1830}}.

\end{thebibliography}

\end{adjustwidth}


%



\end{document}